\documentclass[12pt,preprint]{aastex}

\usepackage{epsfig}
\usepackage{graphicx,color}

\shorttitle{}
\shortauthors{}

\newcommand{\beq}{\begin{equation}}
\newcommand{\eeq}{\end{equation}}
\newcommand{\erg}{{\rm erg}}
\newcommand{\MeV}{{\rm MeV}}
\newcommand{\PeV}{{\rm PeV}}
\newcommand{\zbar}{\bar{z}}
\newcommand{\ybar}{\bar{y}}
\newcommand{\ymax}{y_{\rm max}}

\shorttitle{Reconnection in the Crab Nebula}
\shortauthors{Uzdensky, Cerutti, \& Begelman}

\begin{document}

\title{Reconnection-powered Linear Accelerator and Gamma-Ray Flares in the Crab Nebula}


\author{Dmitri A. Uzdensky and Beno\^it Cerutti}
\affil{CIPS, Physics Department, University of Colorado, 390 UCB, Boulder, CO 80309, USA  
\email {uzdensky@colorado.edu, benoit.cerutti@colorado.edu}} 

\and 

\author{Mitchell C. Begelman 
\affil{JILA, University of Colorado and National Institute of Standards and Technology, 440 UCB, Boulder, CO 80309-0440, USA}
and 
\affil{Department of Astrophysical and Planetary Sciences, University of Colorado,  389 UCB, Boulder, CO 80309-0389, USA
\email {mitch@jila.colorado.edu}}
}



\begin{abstract}

The recent discovery of day-long gamma-ray flares in the Crab Nebula, presumed to be synchrotron emission by PeV ($10^{15}$~eV) electrons in milligauss magnetic fields, presents a strong challenge to particle acceleration models.  The observed photon energies exceed the upper limit ($\sim$100~MeV) obtained by balancing the acceleration rate and synchrotron radiation losses under standard conditions where the electric field is smaller than the magnetic field.  We argue that a linear electric accelerator, operating at magnetic reconnection sites, is able to circumvent this difficulty.  Sufficiently energetic electrons have gyroradii so large that their motion is insensitive to small-scale turbulent structures in the reconnection layer and is controlled only by large-scale fields.  We show that such particles are guided into the reconnection layer by the reversing magnetic field as they are accelerated by the reconnection electric field.  As these electrons become confined within the current sheet, they experience a decreasing perpendicular magnetic field that may drop below the accelerating electric field. This enables them to reach higher energies before suffering radiation losses and hence to emit synchrotron radiation in excess of the 100 MeV limit, providing a natural resolution to the Crab gamma-ray flare paradox. 
\end{abstract}



\date{\today}

\maketitle


\section{Introduction}
\label{sec-intro}

Recently discovered high-energy ($\geq 100\, \MeV$) gamma-ray flares in the Crab Nebula~\citep{AGILE-2011, FERMI-2011} challenge theoretical models of particle acceleration and radiation~\citep{Lyutikov-2010,Komissarov_Lyutikov-2010,Bednarek_Idec-2010}. These flares are believed to come from synchrotron radiation in the nebula, i.e., outside the pulsar wind termination shock, at $> 10^{16}\, {\rm cm}$ \citep{Rees_Gunn-1974,Kennel_Coroniti-1984a}, and not from the vicinity of the pulsar itself. Their most puzzling aspect is the very high photon energies, reaching a few hundred MeV. This exceeds the accepted upper limit (radiation-reaction limit) for photons produced via the synchrotron mechanism, $\epsilon_{\rm sync,c} \simeq (9/4) m_e c^2/\alpha_{\rm fs} \simeq 160\, \MeV$ (where $\alpha_{\rm fs}=e^2/\hbar c$ is the fine structure constant), obtained  by balancing the accelerating electric force on a particle $eE$ with the synchrotron radiation reaction drag force, $f_{\rm rad} \simeq P_{\rm sync}/c = 2 \sigma_T  \gamma^2 B_\perp^2/8\pi$ (where $\gamma$ is the Lorentz factor of the particle, $\sigma_T\equiv 8\pi e^4/3 m_e^2 c^4$ is the Thomson cross-section and ``$\perp$'' means perpendicular to the particle's motion), in combination with the requirement $E< B_\perp$. The latter condition is required in most established particle acceleration mechanisms, e.g., diffusive shock acceleration; thus, these mechanisms cannot accelerate particles to energies high enough to produce the observed synchrotron radiation~\citep{Guilbert_etal-1983,deJager_etal-1996}.

A possible resolution of this paradox is to invoke relativistic Doppler boosting of the emitting region towards the observer \citep[e.g., ][]{Lyutikov-2010,Komissarov_Lyutikov-2010,Bednarek_Idec-2010}. However, although the required bulk Doppler factors ($\sim 3$) are not ruled out, there is little evidence for such high speeds from proper motions in the inner regions of the Crab \citep{Hester_etal-2002} and theoretical arguments suggest that typical bulk speeds downstream of the shock are at best mildly relativistic, unless the shock is strongly oblique \citep[see][]{Komissarov_Lyutikov-2010}.


Here we focus on the alternative possibility that electrons are indeed accelerated to very high (PeV) energies in an extended region where $E>B_\perp$. This is impossible in ideal magnetohydrodynamics (MHD), ${\bf E = - v\times B}/c$, where $v<c$ is the bulk flow velocity, implying $E<B_{\perp}$ (assuming that the particle moves along~${\bf E}$). However, ideal MHD breaks down at magnetic reconnection sites, where $E > B_{\perp}$ is satisfied naturally in a thin layer. Sufficiently energetic particles tend to be drawn into the layer, where they become trapped and may be accelerated to extreme energies, as was suggested by~\cite{Kirk-2004}. By explicitly demonstrating this focusing effect in the ultrarelativistic limit, in this paper we show that a linear accelerator, utilizing the large-scale electric field associated with a reconnection process, can explain gamma-ray flares in the Crab. We describe the mechanism in~\S~2 and its application to the Crab flares in~\S~3.


\section{Extreme Particle Acceleration and Radiation in Reconnecting Current Sheets}
\label{sec-mechanism}
 
\begin{figure}
\epsscale{1.0}
\plotone{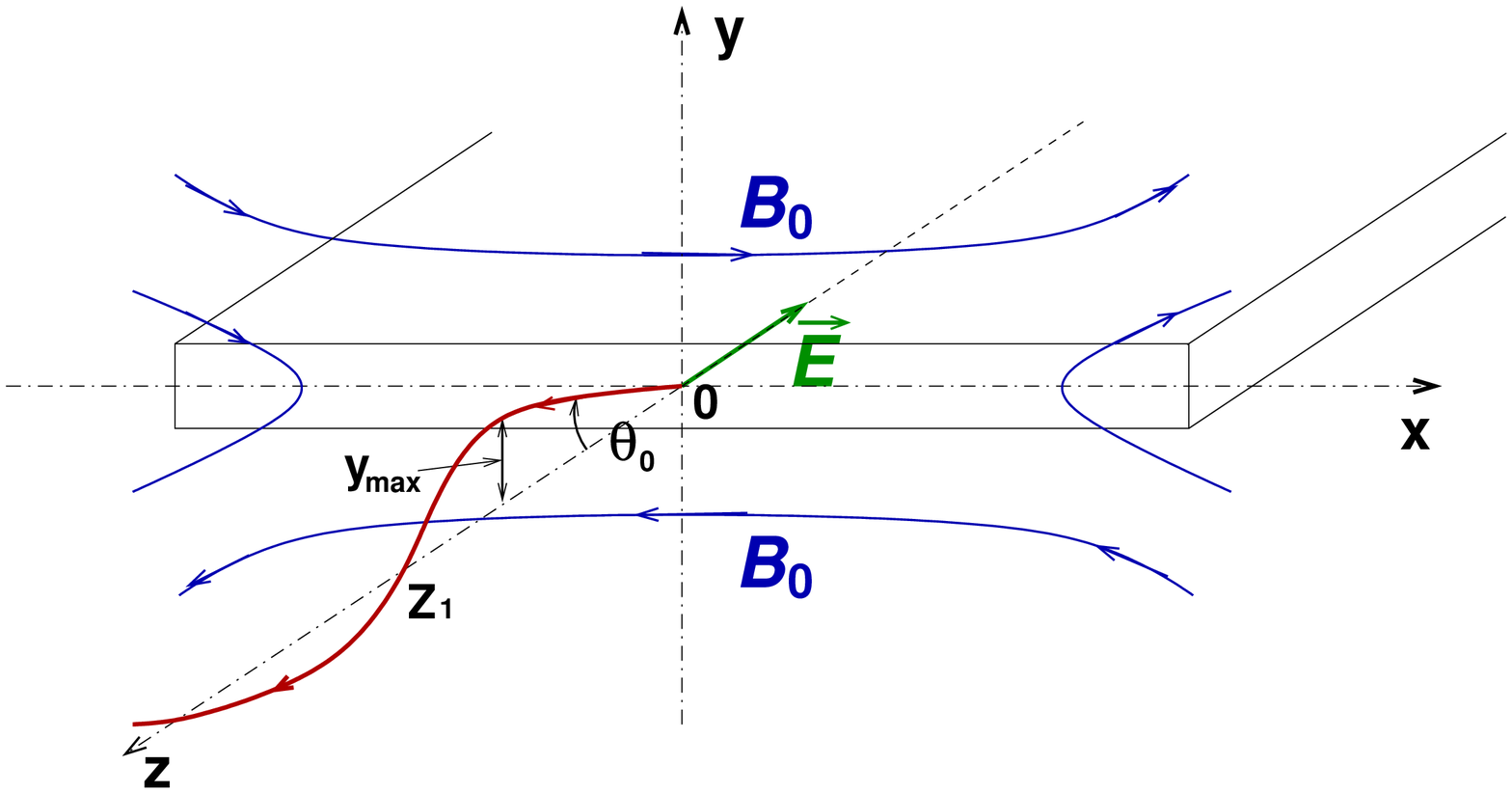}
   \caption{Sketch of a relativistic Speiser orbit of a particle in a reconnection layer.
   \label{fig-1}}
\end{figure}

Reconnection is generally recognized as an important mechanism of non-thermal particle acceleration, including in relativistic pair plasmas thought to exist in the Crab Nebula \citep[e.g.,][]{Zenitani_Hoshino-2001,Zenitani_Hoshino-2008}. Acceleration mechanisms at moderate energies may be complex, e.g., involving small-scale turbulent structures such as multiple magnetic islands~\citep{Drake_etal-2006}. However, the most energetic particles (e.g., the PeV particles considered here) are special: their relativistic gyroradii, $r_L = \gamma\rho_c  \sim 1.7\times 10^{15}\, {\rm cm}\, B_{-3}^{-1}\, \gamma_9$ (where $B_{-3} \equiv B/ 10^{-3}$~G,  $\gamma_9 \equiv \gamma/10^9$,  and $\rho_c \equiv c/\omega_c = m_e c^2/eB \simeq 17\,{\rm km}~B_{-3}^{-1}$) are large, comparable to the global flare region size~$l$ ($<$ 1 light day $\simeq 3\times 10^{15}\,{\rm cm}$ based on observed flare duration). Therefore, small-scale turbulence is washed out for these particles and the complicated mechanisms operating at moderate energies are not important for them. Their motion is sensitive only to the large-scale electromagnetic field structure (Figure~\ref{fig-1}), which consists of the reconnecting magnetic field~$B_x(y)$, reversing across a current layer of some thickness~$\delta$; the nearly-uniform reconnection electric field~$E_z$; and perhaps a guide magnetic field~$B_z$, also roughly uniform. A finite $B_z$ may in fact be required for acceleration of moderate-energy particles in relativistic pair reconnection~\citep{Zenitani_Hoshino-2008}, a pre-condition for injecting seed particles for the highest-energy acceleration considered in this paper. We neglect the reconnected magnetic field~$B_y$. (Note that the Hall magnetic and electric fields are absent in pair plasmas.)

In this electromagnetic field, ultra-relativistic particles move mainly in the $z$-direction in a relativistic analog of Speiser orbits \citep{Speiser-1965,Zenitani_Hoshino-2001,Kirk-2004}. They are accelerated by the electric field~$E_z$ in the $z$-direction while being confined to the layer midplane ($y = 0$) by the reversing reconnecting magnetic field~$B_x$. In the $x$-direction, they may be confined by the guide magnetic field. Importantly, as the particles get accelerated, their orbits not only stretch in the $z$-direction, but also shrink in the $y$-direction; both the Speiser meandering width $y_{\rm max}$ (the maximum deviation from the midplane) and the midplane-crossing angle $\theta_0$ decrease as their energy increases. Thus, the population of energetic particles, even if isotropic initially, gets focused into a tight beam along the current sheet.

We explain this focusing mechanism with the following analytical model. For simplicity, assume $B_z=0$ and consider an electron whose trajectory lies entirely in the $(yz)$ plane.  We choose the starting point at the origin (0,0,0), so that at $z=0$ the particle is just crossing the midplane $y=0$ at some angle $\theta_0< \pi/2$ (see Figure~\ref{fig-1}). We also assume that the electron is already pre-accelerated to a very high Lorentz factor $\gamma_0\gg 1$ and that~$\theta_0$ is not too small, so that both its gyroradius and $y_{\rm max}$ are much larger than the current layer thickness~$\delta$; then, most of the trajectory lies outside the layer, in the two upstream regions with $|B_x(y)| \approx B_0 = {\rm const}$. The electric field is uniform, $E_z = -\, E_0 = - \, \beta_{\rm rec} \, B_0 \simeq -\, 0.3\, {\rm V/cm}\, \beta_{\rm rec}\,B_{0,-3}$, where the dimensionless reconnection rate $\beta_{\rm rec} < 1$ is $\sim 0.1$, as indicated by relativistic particle simulations \citep{Zenitani_Hoshino-2008}. Finally, we take the initial particle energy to be well below the radiation reaction limit.

To determine the evolution of the orbit's key parameters, first consider one half-cycle segment of the orbit, $z\in[0,z_1]$, where $z_1$ is the next midplane-crossing. The motion is determined by two parameters: $\gamma_0$ and the midplane-crossing angle~$\theta_0$.  We wish to calculate the slight changes in $\gamma$ and $|\theta|$ over this segment, i.e., the values $\gamma(z_1)$ and~$|\theta(z_1)|$; from this we will derive the secular evolution of~$\ymax$.

The $\gamma$-factor follows from energy conservation: $\epsilon(z) = \epsilon_0 + eE_0 z$, i.e., 
\beq
\gamma(z) = \gamma_0 + \beta_{\rm rec} \zbar \, , 
\label{eq-gamma-z}
\eeq
where $\zbar \equiv z/\rho_c$. 
The particle's trajectory is obtained from the $y$ component of the relativistic equation of motion: 
$d(\gamma v_y)/dt = - \omega_c v_z\Rightarrow d(\gamma v_y)/dz = 
- \omega_c \equiv - eB_0/m_e c$, 
integrating which we get
\beq
\gamma(\zbar)\, \beta_y(\zbar) =  \gamma_0 \beta_{y0} - \zbar \simeq \gamma_0\sin\theta_0 - \zbar\, ,
\label{eq-gamma-beta_y}
\eeq
where $\beta_{y0} = \beta\, \sin\theta_0 \simeq \sin\theta_0$ since $\beta = v/c \approx 1$. 
This gives us the trajectory's apex~$\zbar_a$ (where $v_y=0$ and $y=\ymax$): 
$\zbar_a = \gamma_0\sin\theta_0$. 

Next, we assume that $|\theta| \leq \theta_0 \ll 1$, so that $d\ybar/d\zbar = \beta_y/\beta_z \simeq \beta_y$, where $\ybar \equiv y/\rho_c$. Then, integrating Equation~(\ref{eq-gamma-beta_y}), we obtain an explicit expression for the particle trajectory: 
$\ybar(\zbar)  = -\, \zbar/\beta_{\rm rec} + 
(\gamma_0/\beta_{\rm rec}^2)\, (1+ \beta_{\rm rec} \sin\theta_0)\, \ln (1+ \beta_{\rm rec} \zbar/\gamma_0)$. 
Anticipating the fractional increase of the particle's energy over~$z_1$ to be small, $\beta_{\rm rec}\zbar_1/\gamma_0 \sim \theta_0 \ll 1$, we expand $\ybar(\zbar)$ and 
%
find $\ybar_{\rm max} \equiv \ybar (\zbar_a) \simeq \gamma_0 \theta_0^2/2$ and the segment's length $\zbar_1$ (the distance to the next midplane crossing, defined by $\ybar(\zbar_1) = 0$): 
$\zbar_1 \simeq 2 \gamma_0 \theta_0 + (2/3)\, \gamma_0\beta_{\rm rec} \theta_0^2$.
%
This allows us to estimate the changes in the orbit parameters, $\gamma$ and $|\theta_0|$, from one midplane crossing to the next: 
$\delta \gamma = \gamma(\zbar_1) - \gamma_0 = \beta_{\rm rec}\,\zbar_1 \simeq 2 \beta_{\rm rec}\, \gamma_0 \theta_0 > 0$, 
and  
$\delta |\theta_0|  = |\theta(\zbar_1)| - \theta_0 \simeq -\, (4/3) \, \beta_{\rm rec} \, \theta_0^2 < 0 $.
Thus, as $\gamma$ increases, $|\theta_0|$ {\it decreases}, i.e., the trajectory becomes increasingly aligned with the  accelerating electric field. 

The secular evolution of $\gamma$ and $|\theta_0|$ over many current-sheet crossing cycles ($z\gg z_1$) follows from $d|\theta_0|/d\gamma \simeq \delta |\theta_0|/\delta\gamma \simeq -(2/3)\,  |\theta_0| /\gamma $, integrating which we get $|\theta_0| \sim \gamma^{-2/3} \sim \zbar^{-2/3}$. Similarly, 
we find $\zbar_1\sim \gamma^{1/3} \sim \zbar^{1/3}$ and $\ymax \sim \gamma^{-1/3} \sim \zbar^{-1/3}$. Thus, the highest-energy particles are focused into a narrow beam confined closer and closer to the midplane! (This shrinkage of the trajectory can be interpreted as a result of the ${\bf E\times B}$ drift of the particle's virtual guiding center away from the midplane, separately in each half-cycle segment. It can also be interpreted as a result of the conservation of the adiabatic invariant $J_y = \int p_y dy \sim \gamma m_e c\, \theta_0 \,\ymax \propto \gamma^2 \, \theta_0^3={\rm const}$.)


Two factors limit the particle energy. First, the finite current-sheet length, $l=10^{16 }l_{16}$ cm, limits the energy to 
\beq
\epsilon_{\rm max} = \gamma_{\rm max} \, m_e c^2 =  e E_0 l = e \beta_{\rm rec} B_0\, l  
\simeq 3\, \PeV \, \beta_{\rm rec} \, B_{0,-3} \, l_{16} \, ,
\label{eq-gamma_max}
\eeq
corresponding to 
$\gamma_{\rm max}  = $ 
$6 \times 10^9  \, \beta_{\rm rec} \, B_{0,-3} \, l_{16}$ and 
\beq
\epsilon_{\rm sync,max} = (3/2) \, \gamma_{\rm max}^2 \, \hbar \omega_c =
(3/2) \,(l/\rho_c)^2\, \hbar\omega_c \simeq 600\, \MeV\, \beta_{\rm rec}^2\, B_{0,-3}^3\, l_{16}^2\, .
\label{eq-epsilon_sync_max}
\eeq

Second, radiation reaction may cause the energy to saturate at a lower value. 
If the initial injection values $y_{\rm max, inj}$ and~$\gamma_{\rm inj}$ satisfy $y_{\rm max,inj}/\delta > (\gamma_{\rm rad,*}/\gamma_{\rm inj})^{1/3}$ (where $\gamma_{\rm rad,*}$ is defined in Equation~(\ref{eq-gamma_rad0}) below), then the radiation reaction-limited regime is reached while most of the orbit is outside the layer, where $|B|= B_0$. 
The cycle-averaged radiation reaction force~$f_{\rm rad}$  
balances the electric force 
when
\beq
\gamma=\gamma_{\rm rad, *} = \sqrt{{3c\over{2r_e\omega_c}}\, \beta_{\rm rec}} \simeq 3\times 10^9\, \beta_{\rm rec}^{1/2}\,B_{0,-3}^{-1/2}\, .
\label{eq-gamma_rad0}
\eeq
By comparing $\gamma_{\rm max}$ and $\gamma_{\rm rad,*}$, we see that the 
radiation reaction limit is  reached before $\gamma_{\rm max}$ only if the layer is sufficiently long, $l_{16} > 0.5\beta_{\rm rec}^{-1/2}\,B_{0,-3}^{-3/2}$. 
Then, the particle moves in a sinusoidal-like orbit with $\gamma\simeq \gamma_{\rm rad,0} ={\rm const}$ and efficiently converts all the energy it gains from the electric field into synchrotron photons with a characteristic energy 
\beq
\epsilon_{\rm sync,*} = (3/2) \, \gamma_{\rm rad,*}^2\, \hbar \omega_c = 
(9/4)\, \beta_{\rm rec}\, \alpha_{\rm fs}^{-1} \, m_e c^2 = 160\, \MeV \,\beta_{\rm rec}\, ,
\label{eq-epsilon_sync*}
\eeq
somewhat too low to explain the Crab flares.

However, despite the saturation of $\gamma$, the particle orbit continues to shrink towards the midplane with $\theta_0\sim \exp(-\beta_{\rm rec}z/3\rho_c \gamma_{\rm rad,*})$. If the layer is long enough, then $\ymax$ eventually shrinks below~$\delta$, reducing radiative losses and thus making extremely high particle energies possible. Indeed, when the entire trajectory is contained deep within the layer, the maximum perpendicular magnetic field sampled by the particle, $B_{\rm max}=B_x(y_{\rm max}) \sim B_0\, y_{\rm max}/\delta$, becomes smaller than~$B_0$, thus reducing~$f_{\rm rad}$ for a fixed~$\gamma$, whereas the accelerating force $eE_0 = e \beta_{\rm rec}\, B_0$ remains unchanged. Correspondingly, the radiation reaction limit $\gamma_{\rm rad}$ increases in compensation to maintain the balance between the cycle-averaged $f_{\rm rad}$ and the constant electric acceleration: $\gamma_{\rm rad} \approx \gamma_{\rm rad,*}\, \delta/\ymax$, and may rise well above~$\gamma_{\rm rad,*}$ as $y_{\rm max}$ shrinks. The corresponding synchrotron photon energy, $\epsilon_{\rm sync} \simeq \epsilon_{\rm sync,*} \, \delta/\ymax$, can easily exceed $\epsilon_{\rm sync,*}$, and thus can explain the $> 100~\MeV$ photon energies observed in the Crab flares. 

We can derive a relationship between $\theta_0$ and~$\gamma_{\rm rad}$ describing their joint evolution in this regime. As a rough estimate, we can regard $B_{\rm max}$ as playing the same role as $B_0$ played for a particle for which most of the trajectory lied outside the layer. Then we can estimate the meandering width as 
$\ymax \simeq 0.5\, \gamma_{\rm rad} \, \theta_0^2 \, \rho_c \, B_0/B_{\rm max} \simeq
0.5\, \gamma_{\rm rad} \,\theta_0^2 \, \rho_c\, \delta/\ymax$, 
i.e., $\ymax\simeq \theta_0\, (\gamma_{\rm rad} \rho_c \delta/2)^{1/2}$. 
Combining this with the expression for the radiation-reaction limited Lorentz factor~$\gamma_{\rm rad}\approx \gamma_{\rm rad,*} \delta/\ymax$ where $\gamma_{\rm rad,*}$ is given by Equation~(\ref{eq-gamma_rad0}), we find
$\theta_0 \simeq \gamma_{\rm rad,*} \, (2\delta/\rho_c)^{1/2}\, \gamma_{\rm rad}^{-3/2}$.

Alternatively, if the initial $y_{\rm max, inj}$ and~$\gamma_{\rm inj}$ are small enough, then the orbit's meandering width $y_{\rm max}$ quickly collapses below~$\delta$, so that the radiation reaction does not become important before the particle reaches the end of the layer. In this case, the particle is simply accelerated by the electric field almost up to the theoretical maximum limit $\gamma_{\rm max}  =  6 \times 10^9 \, \beta_{\rm rec} \, B_{0,-3} \, l_{16}$, which may exceed the conventional radiation-reaction limit given by Equation~(\ref{eq-gamma_rad0}). It will then radiate all its energy in one short (compared to its relativistic cyclotron period $2\pi \gamma\, \omega_c^{-1}$) and powerful burst when it finally escapes the reconnection layer into an adjacent region with a finite magnetic field ($B\sim B_0$), i.e., when it transitions from a big linear accelerator to a powerful radiator (beam dump). If this transition is sharp, then the characteristic synchrotron photon energy will be that given by Equation~(\ref{eq-epsilon_sync_max}).

As ultrarelativistic particles moving on relativistic Speiser orbits cross the layer midplane, they effectively see a rapidly reversing magnetic field, with a characteristic reversal scale $\lambda_B\sim z_1$ much smaller than their gyroradius~$r_L = \gamma\rho_c$. This raises the question of whether the resulting radiation could in fact be jitter radiation~\citep{Medvedev-2000}. However, the jitter parameter~$\delta_{\rm jitter} = \lambda_B/\rho_c$ is much larger than~1 in our case; for example, for particles with $\ymax \gg \delta$,  we have $\lambda_B\sim z_1 \sim 2\gamma_0 \theta_0 \rho_c \gg \rho_c$. Therefore, the radiation emitted by these particles is true synchrotron radiation.

\begin{figure}
\epsscale{1.0}
\plotone{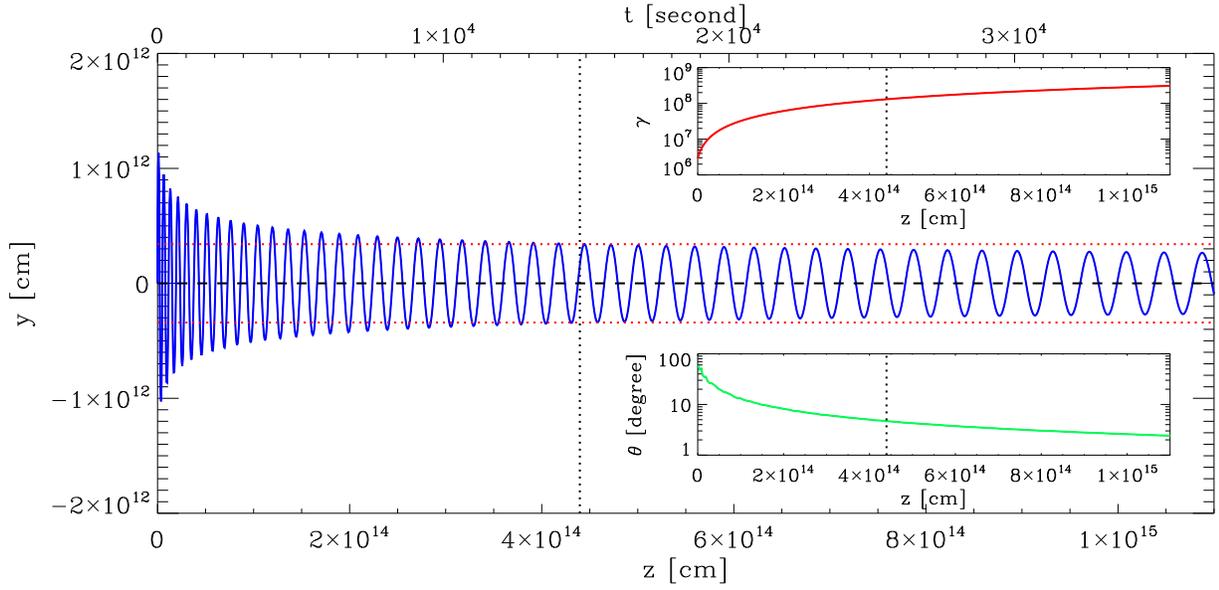}
\caption{Numerically calculated orbit of a relativistic electron in reconnection layer of width $\delta=\gamma_{\rm bulk}\rho_{\rm c}\approx 3.4\times 10^{11}$~cm ($y=\pm\delta$ is shown by red dotted lines), $B_0=5~$mG, $B_{z}=0$, and $\beta_{\rm rec}=0.1$. The particle is initially injected at the origin with $\gamma_{\rm inj}=3\times 10^6$, $\theta_{\rm inj}=90^{\rm o}$. The inserts describe the evolution of the particle's Lorentz factor $\gamma$ and of the midplane-crossing angle~$\theta_0$. The vertical dotted line shows the distance $z$ where the orbit becomes contained in the current layer.
  \label{fig-2}}
\end{figure}
\begin{figure}[htbp]
\epsscale{1.0}
\plotone{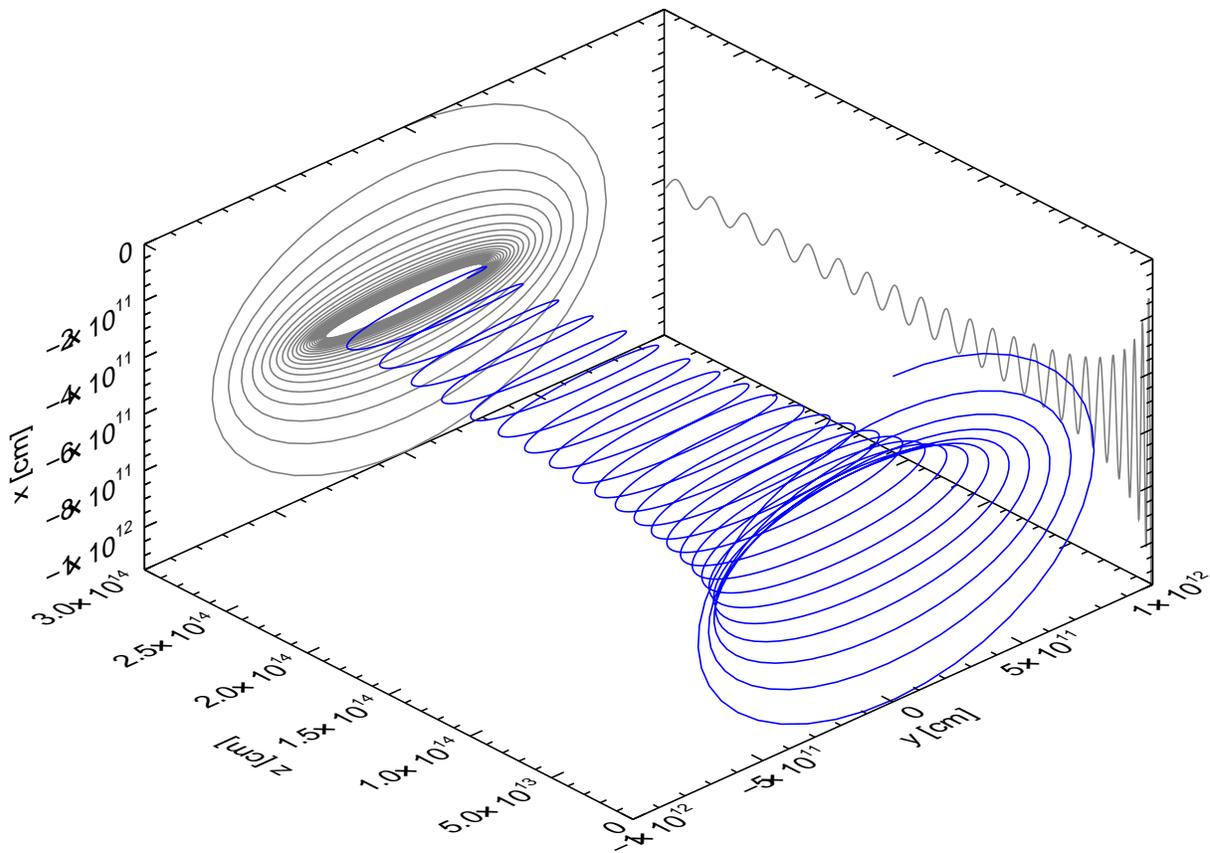}
   \caption{Three-dimensional orbit of a relativistic electron in the reconnection layer with a non-zero guide field $B_{\rm z}=B_0=5~$mG. The other parameters are $\gamma_{\rm inj}=3\times 10^6$, $\theta_{\rm inj}=90^{\rm o}$, $\gamma_{\rm bulk}=10^6$ and $\beta_{\rm rec}=0.1$. The gray lines show the orbit projected onto the $(xy)$ and $(xz)$ planes.}
   \label{fig-3}
\end{figure}
\begin{figure}
\epsscale{1.0}
\plotone{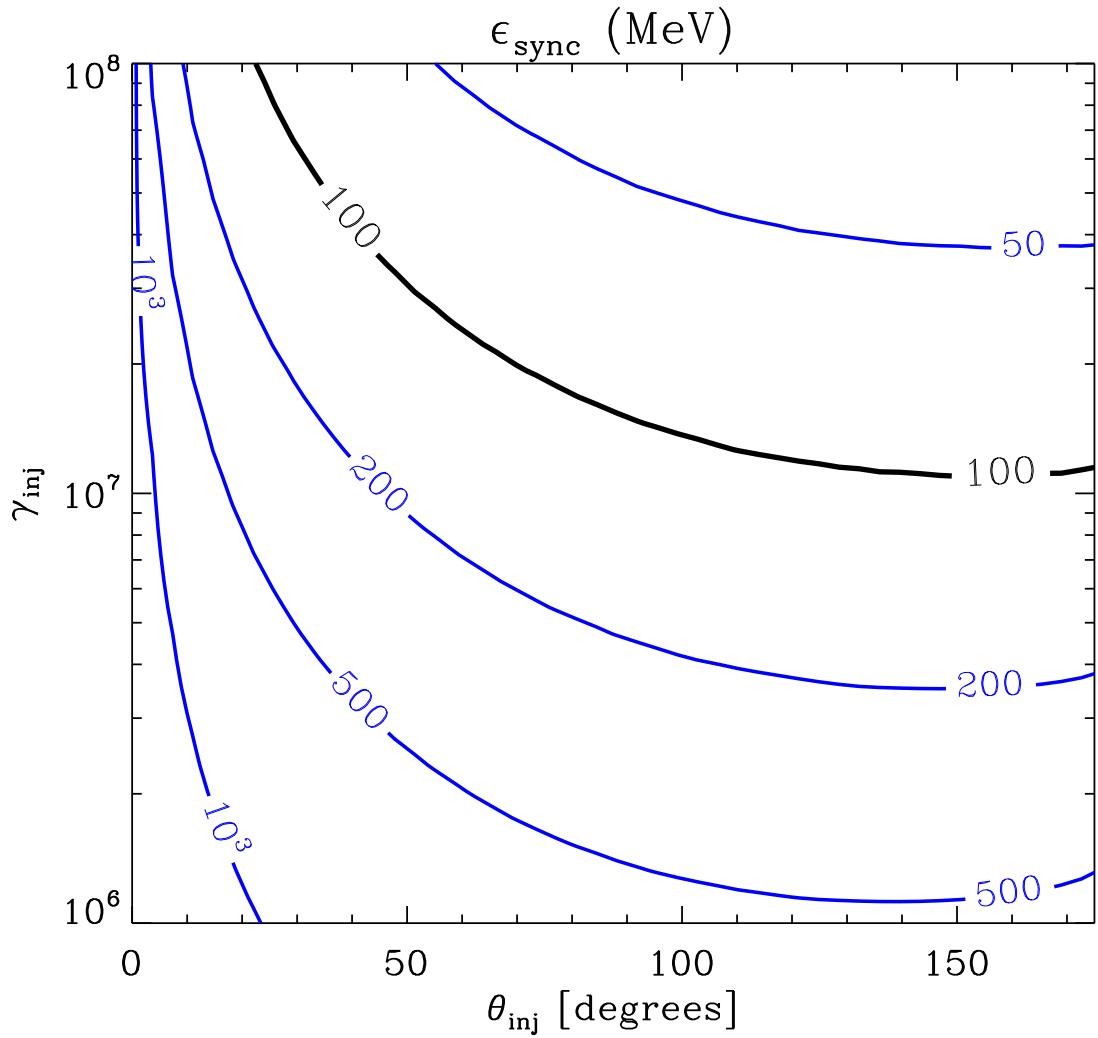}
\caption{Characteristic synchrotron photon energy produced by an electron at the end of the reconnection layer ($l = 4$ light days), as a function of the particle's two initial parameters $\gamma_{\rm inj}$ and~$\theta_{\rm inj}$, with $B_0=5~$mG and $\delta=3.4\times 10^{11}~$cm.
   \label{fig-4}}
\end{figure}

Our analytical model is fully confirmed and generalized by our numerical calculation of 3D relativistic particle orbits  (Figures~\ref{fig-2}-\ref{fig-4}) for arbitrary initial particle parameters, utilizing an explicit 8th-order Runge-Kutta-Verner method \citep{Verner-1978}. 
We include the radiation reaction force,
a guide magnetic field~$B_z$, and a realistic Harris profile of the reconnecting magnetic field, $B_x(y)=B_0\tanh\left(y/\delta\right)$. 
Physically, a lower limit on the current layer thickness~$\delta$ is set by the collisionless skin-depth or the relativistic gyroradius of the bulk electrons: $\delta > \gamma_{\rm bulk} \rho_c = 1.7 \times 10^{12}\, {\rm cm}\, \gamma_{\rm bulk,6} \, B_{-3}^{-1}$, where $\gamma_{\rm bulk,6} = \gamma_{\rm bulk}/10^6$. In reality, however, the layer may be broadened by turbulence driven by secondary instabilities.
 
Our numerical study confirms the focusing effect and the above analytical scaling relationships (Cerutti {\it et al.} 2011, in preparation). 
Figure~\ref{fig-2} shows an electron's orbit in the $(yz)$-plane with no guide field and  
 the evolution of its Lorentz factor and its midplane-crossing angle along the $z$-direction. 
A non-zero guide field adds a circular motion to the particle's orbit in the $(xy)$-plane (see Figure~\ref{fig-3}), 
but the overall evolution of $\gamma$ and $\theta_0$ is almost unchanged even for $B_z=B_0$. 
Figure~\ref{fig-4} gives the characteristic energy $\epsilon_{\rm sync}$ of synchrotron photons radiated by the electrons leaving the layer after four days of acceleration, as a function of the inital parameters 
$\gamma_{\rm inj}$ and $\theta_{\rm inj}$ in a 5~mG magnetic field. 
We find that $>100$~MeV photons can be emitted for $\gamma_{\rm inj}\lesssim 10^7$ 
regardless of the initial angle~$\theta_{\rm inj}$. The particles are confined within a tight cone of 
semi-aperture angle of only a few degrees. 


\section{Application to Crab Gamma-Ray Flares}
\label{sec-crab}

In order to explain the short flare timescales ($\tau_{\rm fl} < 1$~day, see~\citealt{Balbo_etal-2011}), the flaring region must be compact ($l < 10^{16}$ cm), suggesting that the flares originate just outside the pulsar wind termination shock. There are two places where reconnection current sheets may form in the inner Crab Nebula.  First, the equatorial plane contains a reversing toroidal field \citep{Begelman-1999,Komissarov_Lyubarsky-2004}. Second, we expect a cylindrical z-pinch along the rotational axis, which is unstable to kink-like MHD instabilities the nonlinear development of which may lead to episodes of reconnection-driven dissipation (similar to quasi-periodic sawtooth crashes in tokamaks), perhaps powering X-ray emission in the axial jet~\citep{Begelman-1998,Mizuno_etal-2011}. Since the PeV particles are focused along the current sheet, we can see the flare only when the reconnection electric field points toward us.  While neither the equatorial plane nor the rotational axis lie close to our line of sight, instabilities may swing the current sheets through wide angles, allowing us to observe the radiation intermittently. 

Given the compact scale of the flaring region, the voltage drop~$E_0 l$ required to explain the energies of radiating electrons implies $B_0 > $ several~mG. A similar estimate is obtained by requiring $\tau_{\rm fl}$ to exceed the synchrotron cooling time for $>100\,\MeV$-emitting particles (synchrotron cooling can indeed be shown to dominate over adiabatic cooling), yielding $B_{0,-3}^{3/2}\, \tau_{\rm fl,5} > 3$, where $\tau_{\rm fl,5} \equiv \tau_{\rm fl}/10^5\, {\rm s}$. These values are substantially higher than the typical fields in the inner Crab Nebula usually inferred from dynamical arguments \citep{Rees_Gunn-1974,Kennel_Coroniti-1984a}.  But this discrepancy need not be problematic. First, if the axial z-pinch is responsible for the flares, then the field might be amplified by the pinch effect.  Furthermore, the traditional estimates for the field strength in the nebula may be incorrect. Indeed, as is well-established, if ideal MHD holds everywhere outside the termination shock, then the Poynting flux in the wind must be ~$< 1\%$ of the kinetic energy flux  \citep{Rees_Gunn-1974,Kennel_Coroniti-1984a,Begelman_Li-1992}.  This presents a problem, since ultrarelativistic, magnetically-driven winds do not convert most of their magnetic energy to kinetic form \citep{Begelman_Li-1994}.  
Efficient magnetic reconnection downstream of the shock can resolve this problem by dissipating magnetic energy pumped into the nebula by the pulsar wind \citep{Begelman-1998,Lyutikov-2010}. If this is the case, the flares provide an observational clue reconciling the physics of the pulsar wind with the properties of the nebula.  

The strong emission anisotropy in our model, with two oppositely directed beams (one produced by electrons and the other by positrons), can alleviate potential problems with the energetic efficiency of the flares. If the radiation were isotropic, the total energy of $>100$ MeV photons would be about~$4 \times 10^{40}\,\erg$~\citep{AGILE-2011,FERMI-2011}. Since the flare duration limits the emitting region size to about 1 light-day $\simeq 3\times 10^{15}\, {\rm cm}$ \citep{AGILE-2011,FERMI-2011,Balbo_etal-2011}, and hence the volume to about $3\times 10^{46}\, {\rm cm^3}$, the required energy density deposited into PeV electrons (emitting 100~MeV photons) would have to be about $10^{-6}\, {\rm erg/cm^3}$. 
If the energy source for the flare is magnetic (as it is in reconnection), then the required minimum field strength would be $B_{\rm min}  \sim 5\, {\rm mG}\, K^{1/2}$, where $K^{-1} < 1$ is the fraction of the dissipated energy going to the PeV particles.
Thus, even for a magnetic field of several~mG, about 10 times higher than the standard value for the field in the nebula, isotropic emission would require $K\simeq 1$, i.e., a large fraction of the energy to go to PeV particles, which is problematic. 
However, having the emission beamed into two small solid angle cones alleviates this difficulty.

Reconnection-driven particle acceleration may be a persistent mode of dissipation in the Crab. Flaring may occur all the time, but we see powerful gamma-ray flares only rarely, when the emission is beamed towards us. The strongest flares are observed only about twice a year, which is consistent with the typical single flare duration of a couple of days if the beam's solid angle is~$\sim 0.1\,{\rm sr}$.
Although we nominally associate the flare duration with the size of the acceleration region and the synchrotron cooling time for the most energetic electrons, it is also possible that the flare's intrinsic timescale is actually longer but that the current sheet and hence the beam wiggle around. Then, the observed time scale just corresponds to the time for the beam to cross our line of sight and one should expect a nearly symmetric light-curve profile.

This mechanism may also be applicable to other astrophysical systems, e.g., pulsar striped winds and blazar jets, as was first pointed out by \cite{Kirk-2004}.

%

\begin{acknowledgments}
We thank K. Beckwith, R.~Blandford, and R.~Buehler for fruitful discussions. This work was supported by NASA Astrophysics Theory Program grant NNX09AG02G, NSF grant AST-0907872 and NSF grant PHY-0903851.
\end{acknowledgments}


%
%


\end{document}